\documentclass[twocolumn,pre,superscriptaddress]{revtex4}
\usepackage{latexsym}
\usepackage{graphicx}
\usepackage{epsfig,subfigure}
\begin{document}
\title{Power of the power-laws: lessons from unification of small and large time scales for evolution}

\author{Debashish Chowdhury}
\affiliation{Department of Physics, Indian Institute of Technology, Kanpur 208016, India.}

\author{Dietrich Stauffer}
\affiliation{Institute for Theoretical Physics, Cologne University, D-50923 K\"
oln, Euroland.}

\author{Ambarish Kunwar}
\affiliation{Department of Physics, Indian Institute of Technology, Kanpur 208016, India.}

\begin{abstract}
We develop a ``{\it unified}'' model that describes both ``micro'' 
and ``macro'' evolutions within a single theoretical framework. 
The eco-system is described as a {\it dynamic network}; the 
population dynamics at each node of this network describes the 
``micro''-evolution over {\it ecological} time scales (i.e., birth, 
ageing and natural death of individual organisms) while the appearance  
of new nodes, the slow changes of the links and the disappearance 
of existing nodes accounts for the ``macro'' evolution over 
{\it geological} time scales (i.e., the origination, evolution and 
extinction  of species). In contrast to several earlier claims in 
the literature, we observe strong deviations from power law in the 
regime of long life times where the statistics is, usually, poor.

\end{abstract}


\maketitle

The recent surge in the modeling of biological evolution and extinction 
of species, using the concepts and techniques of statistical physics, 
has been stimulated partly by the claims 
(see \cite{drossel,newman,kauffman,bak} for reviews)
that the statistical distributions of several quantities associated 
with the extinction of species follow power laws. However, almost all 
of these models focus only on the ``macro'' evolution (i.e., the 
evolution of species on geological time scales). Neither the birth, 
ageing and, eventually, the death of the individual organisms nor the 
detailed population dynamics make explicit appearance in these 
theoretical descriptions. On the other hand, in reality, a species 
becomes extinct when its entire population is wiped out. 

Therefore, we develop a ``{\it unified}'' model of an eco-system that 
describes both ``micro'' and ``macro'' evolutions. The eco-system is 
described as a dynamic network. The ``micro''-evolution over  
ecological time scales, i.e., birth, growth (ageing) and natural death 
of individual organisms is described by the dynamics within each 
node \cite{staubook}. The network itself evolves slowly with time; over 
sufficiently long time scales populations of some species would drop to 
zero, indicating their extinction, and the corresponding nodes would be 
deleted from the network. On the other hand, appearance of new nodes, 
together with their own population of individual organisms, signals 
origination of new species. In addition, the links of the network also 
change slowly to capture the adaptive evolution of the species by random 
mutations over geological time scales.

\noindent {\it The dynamic network}:
At any arbitrary instant of time $t$ the model consists of $N(t)$ 
{\it species} each of which may be represented by one of the $N$ 
nodes of a dynamic network; the total number of nodes is not 
allowed to exceed $N_{max}$. Our model allows $N(t)$ to fluctuate 
with time over the range $1 \leq N(t) \leq N_{max}$. The population 
(i.e., the total number of organisms) of a given species, say, $i$ 
at any arbitrary instant of time $t$ is given by $n_i(t)$. The 
limited availablity of resources, other than food, in the eco-system 
imposes an upper limit $n_{max}$ of the allowed population of each 
species. Thus, the total number of organisms $n(t)$ at time $t$ is 
given by $n(t) = \sum_{i=1}^{N(t)} n_i(t)$. Both $N_{max}$ and 
$n_{max}$ are time-independent in the model.

\noindent {\it The interactions}:
Prey-predator interactions are captured through the matrix ${\bf J}$. 
The influence {\it of} species $j$ {\it on} species $i$ is given by 
$J_{ij}$; in general, $J_{ij} \neq J_{ji}$. The only restriction we 
impose initially on the elements of ${\bf J}$ is that $J_{ii} = 0$, 
i.e., none of the organisms preys on any other member of the same 
species. Since in all practical situations, food webs specify only 
the sign of $J_{ij}$ we allow the off-diagonal elements of $J_{ij}$ 
to take only the  values $+1$ and $-1$. Thus, $J_{ij}=1=-J_{ji}$ 
indicates that $j$ is the prey and $i$ is the predator. Similarly, 
the situations $J_{ij}=-1=J_{ji}$ and $J_{ij}=1=J_{ji}$ correspond, 
respectively, to {\it competition} and {\it cooperation} between the 
species $i$ and $j$. We assign the values $+1$ or $-1$ to the 
off-diagonal elemts of ${\bf J}$ randomly with equal probability in 
the initial state of the eco-system \cite{sole}. However, our model 
can be easily generalized to take into account any other architecture 
of food webs \cite{amaral,abramson}.

We now argue that the matrix ${\bf J}$ accounts not only for the 
{\it inter}-species interactions but also {\it intra}-species 
interactions. First of all, note that if $J_{ij} > 0$, then the 
species $j$ is a prey of the species $i$ if, simultaneously, 
$J_{ji} < 0$ whereas the species $j$ cooperates with $i$ if, 
simultaneously, $J_{ij} > 0$ and $J_{ji} > 0$. Therefore, 
if $J_{ij} > 0$, the quantity $(J_{ij} - J_{ji})/2$ is unity if 
the species $j$ is a prey of the species $i$, but it vanishes if 
the species $i$ and $j$ mutually cooperate. Similarly, if 
$J_{ij} < 0$, the quantity $-(J_{ij} - J_{ji})/2$ is unity if the 
species $j$ is a predator of $i$, but it vanishes if the species 
$i$ and $j$ compete against each other. Now, consider the two sums 
\begin{equation}
{\cal S}_i^{\pm} = \pm \sum_{j=1}^N \frac{(J_{ij}^{\pm} - J_{ji})}{2} n_j 
\end{equation}
where the superscript $\pm$ on $J_{ij}$ indicates that the sum is 
restricted to only the positive (negative) elements $J_{ij}$. The 
sum ${\cal S}_i^{+}$ is a measure of the total food {\it currently} 
available to the $i$-th species whereas $-{\cal S}_i^{-}$ is a measure 
of the total population of the $i$-th species that would be, at the 
same time, consumed as food by its predators. If the food available 
is less than the requirement, then some organisms of the species $i$ 
will die of {\it starvation}, even if none of them is killed by any 
predator. This way the matrix ${\bf J}$ can account for the shortfall 
in the food supply and the consequent competition among the organisms 
of the species $i$ .

\noindent {\it The collective characteristics of species}:
The age of an arbitrary individual organism, say, $\alpha$ of the 
species $i$ at time $t$ is denoted by the symbol $X(i,\alpha;t)$.
In our model each species $i$ is {\it collectively} characterized 
by \cite{stauffer,sole}:\\
(i) the {\it minimum reproduction age} $X_{rep}(i)$, 
(ii) the {\it birth rate} $M(i)$,
(iii) the {\it maximum possible age} $X_{max}(i)$, and 
(iv) the elements $J_{ij}$ and $J_{ji}$ ~$(j=1,2,...,N)$. 
An individual of the $i$-th species can reproduce only after attaining 
the age $X_{rep}(i)$. Whenever an organism of $i$-th species gives 
birth to offsprings, $M(i)$ of these are born simultaneously. None of 
the individuals of the $i$-th species can live longer than $X_{max}(i)$. 
Thus, even if an individual manages to escape its predators, it 
cannot live longer than $X_{max}(i)$ because of ``natural death'' 
caused by ageing. 

\noindent{\it The dynamics of the eco-system}:
The state of the system is updated in discrete time steps as follows: 

\noindent {\it Step I- Birth}: Assuming, for the sake of simplicity, the 
reproductions to be {\it asexual}, each individual organism $\alpha$ 
($\alpha = 1,...,n_i(t)$) of the species $i$ ($i=1,2,...N(t)$) is 
allowed to give birth to $M(i;t)$ offsprings at every time step $t$ 
with probability (per unit time) $p_b(i,\alpha;t)$ which is non-zero 
only when the individual organism age $X(i,\alpha;t) > X_{rep}(i;t)$. 

\noindent {\it Step II- Natural death}: At any arbitrary time step $t$ the 
probability (per unit time) of ``natural'' death (due to ageing) of 
an individual organism $\alpha$ of species $i$ is $p_d(i,\alpha;t)$. 

\noindent {\it Step III- Mutation}: With probability $p_{mut}$ per unit time, 
all the species simultaneously re-adjust one of the incoming 
interactions $J_{ij}$ by assigning it a new value of either $+1$ or 
$-1$ with equal probability \cite{sole}. 

\noindent {\it Step IV- Starvation death and killing by prey}: 
If $n_i-{\cal S}^{+}$ is larger than ${\cal S}^{-}$ then food 
shortage will be the dominant cause of premature death of a 
fraction of the existing population of the species $i$. On the 
other hand, if ${\cal S}^{-} > n_i-{\cal S}^{+}$, then a fraction 
of the existing population will be wiped out primarily by the 
predators. In order to capture these phenomena, at every time 
step $t$, in addition to the natural death due to ageing, a 
further reduction of the population by  
\begin{equation}
C ~~\max({\cal S}^{-},~n_i-{\cal S}^{+})
\label{eq-kill}
\end{equation}
is implemented where $n_i(t)$ is the population of the species $i$ 
that survives after the natural death step above. $C$ is a constant 
of proportionality that basically sets the time scale of the 
population dynamics. If implementation of these steps makes $n_i \leq 0$, 
species $i$ becomes extinct.  

\noindent {\it Step V- Speciation}: At each time step, the niches 
(nodes) left empty by extinction of species are re-filled by new 
species, with probability $p_{sp}$. All the simultaneously re-filled 
nodes of the network originate from {\it one common ancestor} which 
is picked up randomly from among the surviving species. All the 
interactions $J_{ij}$ and $J_{ji}$ of the new species are identical 
to those of their common ancestor; each new species, however, either 
competes or cooperates with its ancestor species. The characteristic 
parameters $X_{max}$, $X_{rep}$, $M_j$ of each of the new species 
differ randomly by $\pm 1$ from the corresponding parameters for 
their ancestor.  

\noindent{\it Probability of birth}:
We chose the {\it time-dependent} probabilty of birth per unit time as 
\begin{eqnarray}
p_b(i,\alpha) = \biggl[\frac{X_{max}(i) - X(i,\alpha)}{X_{max}(i) - X_{rep}(i)}\biggr]\biggl(1-\frac{n_i}{n_{max}}\biggr) \nonumber \\
{\rm iff}~ X(i,\alpha) ~\geq ~ X_{rep}(i) ~{\rm and}~ X_{max}(i) > X_{rep}(i)
\end{eqnarray}
Note that in the limit of vanishingly small population, i.e., 
$n_i \rightarrow 0$, we have $p_b(i,\alpha) \rightarrow 1$  
if $X(i,\alpha) = X_{rep}(i)$ and, then, $p_b$ decreases linearly  
\cite{austad} as the organism grows older. However, since the eco-system 
can support only a maximum of $n_{max}$ individual organisms of each 
species, $p_b(i,\alpha;t) \rightarrow 0$ as $n_i(t) \rightarrow n_{max}$, 
irrespective of the age of the individual organism $\alpha$ \cite{cebrat}.

\noindent{\it Probability of natural death}:
We assume the probability of ``natural'' death (due to ageing) to 
have the form \cite{carey}
\begin{eqnarray}
p_d(i,\alpha) = \biggl[\frac{X(i,\alpha) M(i) - X_{rep}(i)}{X_{max}(i) M(i) - X_{rep}(i)}\biggr] \nonumber \\
{\rm if}~ X(i,\alpha) ~\geq ~ X_{rep}(i) 
\end{eqnarray}
\begin{eqnarray}
p_d(i,\alpha) = \biggl[\frac{X_{rep}(i) M(i) - X_{rep}(i)}{X_{max}(i) M(i) - X_{rep}(i)}\biggr] \nonumber \\
{\rm if}~ X(i,\alpha) ~< ~ X_{rep}(i)
\end{eqnarray}
provided $X_{max}(i)M(i) > X_{rep}(i)$. In all other situations,
$p_d(i,\alpha)=1$. Note that, for a given $X_{max}$ and $X_{rep}$, 
the larger is the $M$ the higher is the $p_d$ for any age $X$. 
Therefore, each species have a tendency to increase $M$ for giving  
birth to larger number of offsprings whereas the higher mortality 
for higher $M$ opposes this tendency.

\begin{figure}[tb]
\begin{center}
\includegraphics[angle=-90,width=0.9\columnwidth]{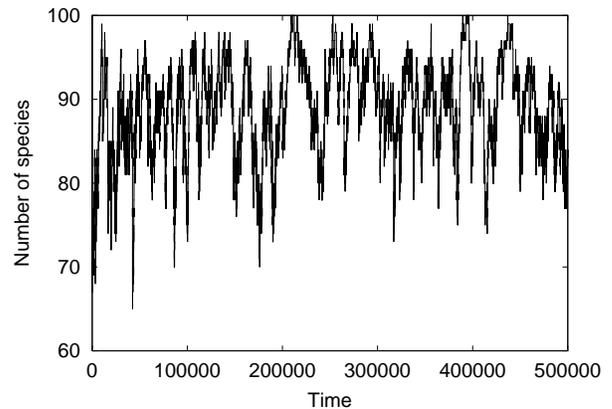}
\end{center}
\caption{The total number of species $N(t)$ is plotted against time; 
the corresponding parameter set is: $N_{max} = 100, n_{max} 1000$, 
$C = 0.1, p_{sp} = 0.001, p_{mut} = 0.001$. 
}
\label{fig-1}
\end{figure}

\begin{figure}[tb]
\begin{center}
\includegraphics[angle=-90,width=0.9\columnwidth]{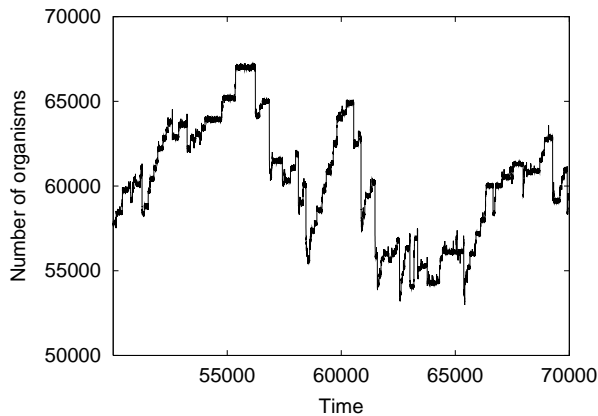}
\end{center}
\caption{The total number of organisms $n(t)$ is plotted against time 
over a {\it relatively short part} of the evolution of the eco-system. 
The parameter values are identical to those in fig.\ref{fig-1}.
}
\label{fig-2}
\end{figure}

The longest runs in our computer simulations were continued upto a 
maximum of {\it five million} time steps. If each time step in our 
model is assumed to correspond to a real time of the order of one 
year, then the time scale of $5$ million years, over which we have 
monitored our model eco-system, is comparable to real geological time 
scales. 

\begin{figure}[tb]
\begin{center}
\includegraphics[angle=-90,width=0.9\columnwidth]{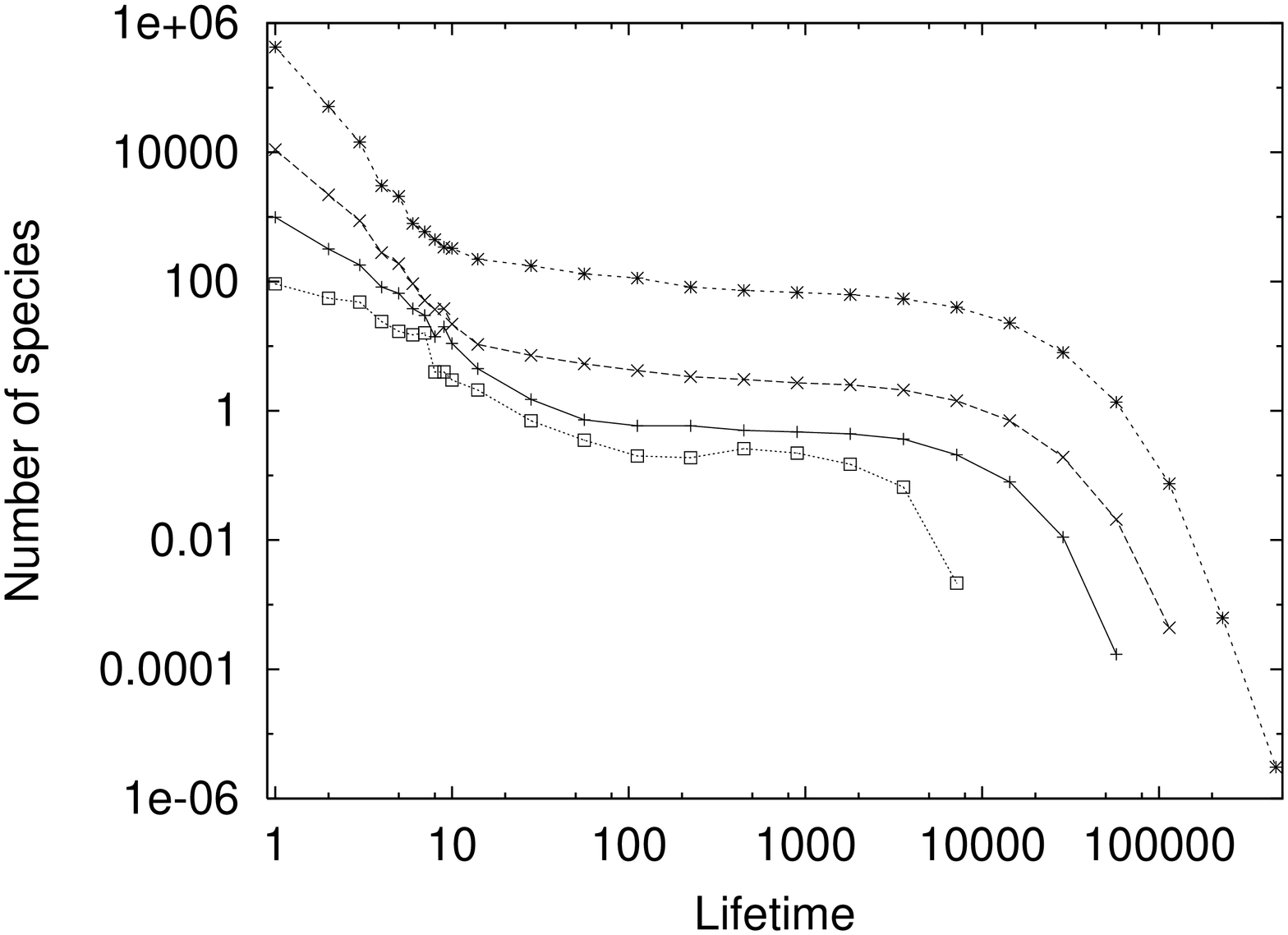}
\includegraphics[angle=-90,width=0.9\columnwidth]{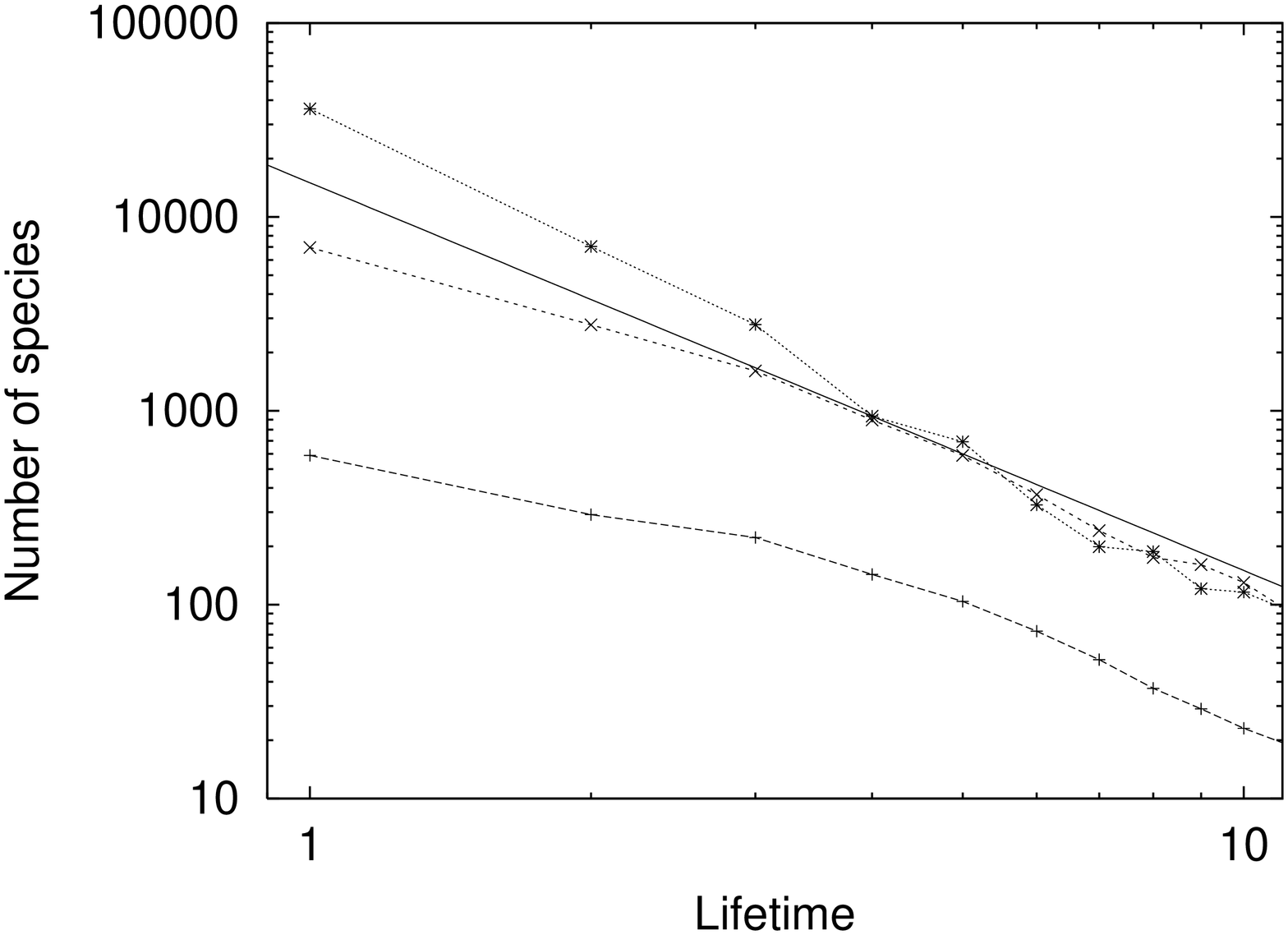}
\end{center}
\caption{Log-log plots of the distributions of the lifetimes of the 
species in an eco-system with (a) $n_{max} = 100$, (b) $n_{max} = 1000$. 
In (b) the data beyond the lifetime of $10$ are not shown to emphasize 
the initial power law regime (symbolized by the straight line with 
slope $-2$ . The other common parameters for both the figures are 
$N_{max} = 50, p_{sp} = 0.01, p_{mut} = 0.001$. The symbols 
$\Box, +, \times$ and $\ast$ in (a) correspond to maximum simulation 
times $5 \times 10^3, 5 \times 10^4, 5 \times 10^5$ and $5 \times 10^6$, 
respectively while the symbols $+, \times$ and $\ast$ in (b) correspond 
to the maximum simulation times $10^4, 5 \times 10^4$ and $4 \times 10^5$, 
respectively.  Each of the data points has been obtained by averaging 
over 18 to 176 runs, each starting from a new random initial state.}
\label{fig-3}
\end{figure}

Since we faced difficulty in getting high quality data, with 
reasonably good statistics, for $N_{max} > 100$ and $n_{max} > 1000$, 
we have carried out most of our simulations with $N_{max} = 50, 100$ 
and $n_{max} = 100, 1000$ only. The data obtained from the different 
runs, each starting from a random initial condition, were averaged. 
Both CRAY-T3E and SUN workstations were used for the simulations.

\begin{figure}[tb]
\begin{center}
\includegraphics[angle=-90,width=0.9\columnwidth]{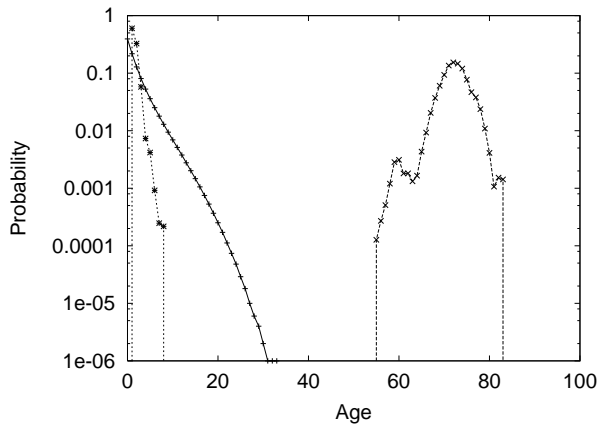}
\end{center}
\caption{Semi-log plot of the distributions of $X$ ~($+$), 
$X_{max}$ ~($\times$) and $X_{rep}$ ~($\ast$); the parameter 
values are same as those in fig.\ref{fig-1}.
}
\label{fig-4}
\end{figure}

In fig.\ref{fig-1} we plot the total number of species, $N(t)$, in 
a particular run, starting from a single initial condition, upto half 
a million time steps. In fig.\ref{fig-2} we plot the corresponding 
variation of the total population $n(t)$ over relatively short 
interval of $20,000$ time steps only. These clearly demonstrate that 
the evolution has periods of ``stasis'' during which organisms 
populations keep fluctuating; the stasis are interrupted by occasional 
bursts of rapid extinctions followed by slower recovery.

The average distributions of the lifetimes of the species are plotted 
in fig.\ref{fig-3} for one set of values of the parameters. Clearly, 
the data are consistent with a power-law; the effective exponent, which 
is, approximately, $2$, is also consistent with the corresponding estimate 
quoted in the literature \cite{drossel,newman}. However, in 
fig.\ref{fig-3} the power law holds only over a limited range \cite{chu}. 
Since real eco-systems  are much more complex than our model eco-system 
and the available fossil data are quite sparse, it is questionable 
whether real extinctions follow power laws and, if so, over how many 
orders of magnitude.

\begin{figure}[tb]
\begin{center}
\includegraphics[angle=-90,width=0.9\columnwidth]{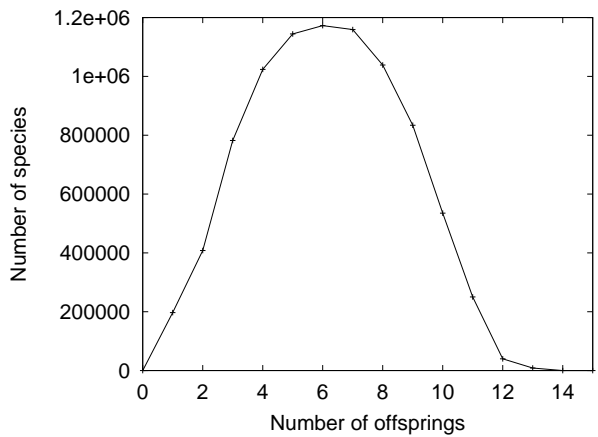}
\end{center}
\caption{The distribution of $M$; the parameter 
values are same as those in fig.\ref{fig-1}.
}
\label{fig-5}
\end{figure}

We have also observed (not 
shown) that the higher is the mutation probability $p_{mut}$ the lower 
is the lifetime; this is consistent with the intuitive expectation that 
the higher rate of mutation leads to higher levels of biological activity 
in the eco-system thereby leading to the extinction of larger number of 
species. But, $p_{sp}$ had weaker effect on the same data. However, 
if $p_{sp}$ is too small to maintain adequate pace of recovery of the 
eco-system after mass extinctions, the entire eco-system collapses.

Fig.\ref{fig-4} shows the time-averaged age-distribution in  the 
populations of a species as well as the distributions of $X_{max}$, 
$X_{rep}$ and $M$. We see that the minimum age of reproduction 
$X_{rep}$ is quite small, as usual in the employed ageing model 
\cite{stauffer}. The age distribution decays stronger than a simple 
exponential, indicating a mortality increasing with age as it should 
be \cite{carey}. The genetic death ages $X_{max} \simeq 72$ are far 
above the upper end $\simeq 31$ of the age distribution, as is
appropriate for animals in the wild \cite{austad}. Finally, 
fig.\ref{fig-5} shows the distribution of $M(i)$; this is relatively 
much broader than the distributions of $X_{max}$ and $X_{rep}$.

In this letter we have presented a unified model which describes 
the birth, ageing and death of individuals and population dynamics 
on short time scales as well as the long-time evolution of species. 
Not only the total number of species and the inter-species 
interactions but also the collective characteristics, namely, 
$X_{rep}, X_{max}$ and $M$, of each species vary following a 
stochastic dynamics. Thus, our model is capable of 
{\it self-organization}. To our knowledge, there are only a few 
earlier evolutionary models \cite{abramson} based on inter-species 
interactions which describe population dynamics of each species. 
The population dynamics  within the framework of Lotka-Volterra 
equations have been considered earlier \cite{jan} for only a few 
species. But, these do not account for the age distributions as the 
entire population of each species is represented {\it collectively} 
by a single dynamical variable in contrast to the explicit birth, 
growth and death of individual organisms captured in our model. 

Since we have observed strong deviations from power-law in the 
distributions of lifetimes of species even in the Sole-Manrubia 
model \cite{chowstau}, in spite of the absence of detailed 
"micro"-dynamics in the latter, we strongly believe that this 
is a generic features of evolution and extinction of species. 
It would be interesting to investigate the {\it geographical effects} 
on our model eco-system by re-formulating it on a lattice in the 
same spirit in which some lattice models of prey-predator systems 
have been formulated \cite{tome}.

We thank P.M.C. de Oliveira for comments on the manuscript and the 
Supercomputer Center J\"ulich for computer time on their CRAY-T3E.
This work is supported by Deutsche Forschungsgemeinschaft through a 
Indo-German joint research project and by Alexander von Humboldt 
Foundation.  

\bibliographystyle{plain}

\end{document}